\documentclass[twocolumn,english,aps,graphicx,amsmath,showpacs]{revtex4}
\usepackage[T1]{fontenc}
\usepackage[latin1]{inputenc}
\usepackage{babel}
\usepackage{graphics}

\makeatletter

\providecommand{\LyX}{L\kern-.1667em\lower.25em\hbox{Y}\kern-.125emX\@}

\newcommand{\J}{{\mathbf {w}}}    \newcommand{\x}{{\mathbf {x}}} \newcommand{\W}{{\mathbf {w}}}         

\makeatother
\begin{document}

\title{Mutual learning in a tree parity machine and its application to cryptography}

\author{Michal Rosen-Zvi\( ^{1} \), Einat Klein\( ^{1} \), Ido Kanter\( ^{1} \)
and Wolfgang Kinzel\( ^{2} \)}

\affiliation{\( ^{1} \)Minerva Center and Department of Physics, Bar-Ilan University,
Ramat-Gan, 52900 Israel,}

\affiliation{\( ^{2} \)Institut f\"ur Theoretische Physik, Universit\"at W\"urzbur,
Am Hubland 97074 W\"urzbur, Germany}

\begin{abstract}
Mutual learning of a pair of tree parity machines with continuous
and discrete weight vectors is studied analytically. The analysis
is based on a mapping procedure that maps the mutual learning in tree
parity machines onto mutual learning in noisy perceptrons. The stationary
solution of the mutual learning in the case of continuous tree parity
machines depends on the learning rate where a phase transition from
partial to full synchronization is observed. In the discrete case
the learning process is based on a finite increment and a full synchronized
state is achieved in a finite number of steps. The synchronization
of discrete parity machines is introduced in order to construct an
ephemeral key-exchange protocol. The dynamic learning of a third tree
parity machine (an attacker) that tries to imitate one of the two
machines while the two still update their weight vectors is also analyzed.
In particular, the synchronization times of the naive attacker and
the flipping attacker recently introduced in \cite{Shamir} are analyzed.
All analytical results are found to be in good agreement with simulation
results.
\end{abstract}


\maketitle

\section{Introduction}

Artificial neural networks are known for their ability to learn \cite{Krogh,EnvB}.
They produce an output from a given input according to some weight
vector and a transfer function. Traditionally, there are two types
of learning. One type is unsupervised learning where a network receives
input and tries to learn about the input distribution. The other type
is the teacher-student scenario, when the so-called teacher receives
inputs, produces outputs and gives another machine, the so-called
student, both the inputs and their assigned outputs. In such a scenario
the teacher is static, i.e., its weight vector does not change during
the learning, and the student tries to imitate the teacher so as to
produce the same output in a new unknown example by dynamically updating
its weight vector. The state in which the student achieves the same
weight vector as that of the teacher and can therefore perform the
same output as that of the teacher is referred to as perfect learning. 

During the last few years a new type of learning scenario has been
introduced and is under discussion: the \emph{mutual} \emph{learning}
procedure. In the mutual learning procedure there is no distinction
between the teacher role and the student role; both networks function
the same way. They receive inputs, calculate the outputs and update
their weight vector according to the match between their mutual outputs
\cite{MetzlerKinzelKanter,Kinzel}. This is an online learning procedure
where in each step one input vector is given, the output in both machines
is calculated and the resulting increment of each weight vector is
added accordingly. It was found that perceptrons that undergo \emph{mutual
learning} might end up in a synchronized state when the weight vectors
of both machines are either parallel - exactly the same, or anti-parallel
- exactly the opposite (depending on their specific updating rule).
The stationary synchronized solution is equivalent to the stationary
perfect learning solution in the teacher-student scenario. We extend
the analysis of mutual learning between perceptrons to mutual learning
between parity machines . We introduce a generic method of analyzing
mutual learning in feedforward tree multi-layer networks where we
concentrate on the tree parity machine (TPM)\cite{pm1,pm2,pm3}. The
method is based on a mapping procedure that maps the mutual learning
in TPMs onto mutual learning in noisy perceptrons. 

A novel cryptosystem composed of two parity machines that synchronize
has recently attracted much attention \cite{KKK,Shamir,RKK,Mislovaty}.
A host of simulation results show that discrete TPMs can synchronize
very fast and a third machine that tries to learn their weight vector
achieves only partial success. These properties make mutual learning
in TPMs attractive for applications in secure communications, as an
information-bearing message can be hidden within a complicated structure
of the TPM's weight vectors and still be reconstructed at the receiver
using another TPM whose parameters are exactly matched to those of
the first one. This type of cryptosystem can provide a new basis for
security much different from currently used cryptosystems that involve
large integers and are based upon number theory \cite{Stinson}. 

The discrete machines studied carried out an updating procedure different
from the conventional learning procedures analyzed in neural networks.
In the discrete machine procedure the increment of the weight vector
in each step is finite and not infinitesimally small. Since the methods
of analyzing discrete on-line learning in contemporary research, see
\cite{VdBr,Schiest,KinzelU,RosenZvi,RosenKanter}, are not applicable
to this case, we introduce here a novel method for analyzing mutual
learning in networks with discrete weight vectors and a learning process
that is based on a finite increment. First, we describe mutual learning
with discrete \emph{perceptrons}, and then we exploit the method of
mapping mutual learning between TPMs onto mutual learning between
noisy perceptrons and analyze mutual learning in discrete TPMs. 

In cryptography, one of the most important aspects of the channel
is its security. Therefore, potential algorithms of eavesdroppers
are included in our analysis. Such algorithms are actually sophisticated
learning procedures where the parties are the teachers and their weights
are time dependent, and the eavesdropper is the student. In the following
we name this time-dependent-teacher-student scenario \emph{dynamic
learning}.

In this Paper we analyze mutual learning and dynamic learning in TPMs
of two kinds: machines with continuous weight vectors (the spherical
constraint - see Eq. (\ref{sphere}) below) and with discrete weight
vectors and finite increment (see Eq. (\ref{discretetry2}) below).
We introduce a method that maps mutual learning in two layered parity
machines onto mutual learning in noisy perceptrons. The spherical
tree parity machine is studied using the same tool box used for studying
mutual learning in the perceptron \cite{MetzlerKinzelKanter}. The
interesting behavior of full synchronization for a certain regime
in the learning rate space and partial synchronization in the other
regime is also found in the mutual learning of TPMs. Mutual learning
in a TPM when the weight vectors are continuous is described by equations
of motion that reveal the evolution of the order parameters in time.
The derivation of the equations of motion is based on the assumption
that the order parameters are self-averaging quantities \cite{kadanoff,Reens}.
This assumption is violated when the increment of the weight vectors
in each step is finite and not infinitesimally small, as in the case
of the discrete weight vector studied here. Therefore we develop different
analytical tools for the case of discrete weight vectors. 

This Paper is an extension of \cite{RKK}. It contains a full, detailed
description of the analytical methods and discussions that were not
included in \cite{RKK}. An advanced attack suggested recently by
Shamir et al \cite{Shamir} - the flipping attack - is also analyzed.
The paper is organized as follows: in section \ref{model} we introduce
the TPM model. We employ a general framework to present its application
to Cryptography in \ref{generalFW}. The dynamics studied are presented
in \ref{dynamic} and the order parameters and local field distributions
are discussed in \ref{OrderP}. The mapping procedure is detailed
in \ref{mapping}. The learning in continuous TPMs is given in \ref{CPM},
where we divided the section into mutual learning (section \ref{antipa}),
and dynamic learning (section \ref{dylear}). The section is summarized
and the results are discussed in \ref{summcont}. Discrete learning
is presented in section \ref{DM}. We first describe mutual learning
in perceptrons in \ref{discPerc}. The extension to mutual learning
in parity machines is given in \ref{discPM}. Two dynamic learning
attacks are studied, the naive attacker (in \ref{simpleA}), and the
flipping attacker (in \ref{Geometric}). A discussion and an overview
are given in \ref{DMdiscus}. All analytical results are found to
be in good agreement with simulation results as indicated in each
section.

\section{\label{model}The model}

\begin{figure}
{\centering \resizebox*{0.5\textwidth}{0.3\textheight}{\includegraphics{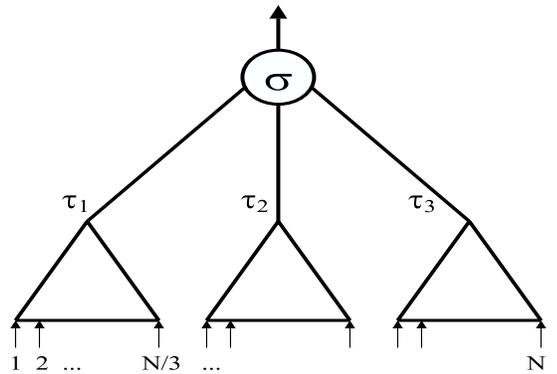}} \par}

\caption{\label{TPMfig}A tree parity machine \protect\( N:3:1\protect \)}
\end{figure}

We consider a TPM with \( K \) binary hidden units \( \tau _{i}=\pm 1,\quad i=1,...,K \)
feeding a binary output, \( \sigma =\prod _{i=1}^{K}\tau _{i} \),
see Figure \ref{TPMfig}. The networks consist of either a continuous
or a discrete coupling vector \( \W _{i}=W_{1i},...,W_{Ni} \) and
disjointed sets of inputs \( \x _{i}=X_{1i},...,X_{Ni} \) containing
\( N \) elements each. The input elements are random variables with
zero mean and unit variance. We confine the input components to \( x_{ji}=\pm 1 \)
without losing generality. The local field in the \( i \)th hidden
unit is defined as\begin{equation}
\label{local}
h_{i}=\frac{1}{\sqrt{N/3}}\W _{i}\x _{i,}
\end{equation}
and the output in the \( i \)th hidden unit is derived by taking
the sign of the local field. The output of the tree parity machine
is therefore given by \[
\sigma =\prod _{i=1}^{K}sign(h_{i})=\prod _{i=1}^{K}\tau _{i}.\]
Our analysis is limited to TPMs with three hidden units, \( K=3 \),
merely for simplicity of the representation of the analysis. The extension
of the formalism to any number of hidden units is straightforward.

The weight vectors of the TPMs are initiated at random according to
a certain constraint. We studied two different cases: the case when
the weight vectors are confined to a sphere, \begin{equation}
\label{sphere}
\sum _{j=1}^{N}W_{ji}^{2}=N,
\end{equation}
 and are initiated randomly according to a Gaussian distribution;
and the case when there are a finite number of available integer values
that each component of the weight vector can take, \begin{equation}
\label{discretetry2}
W_{ji}=\pm L,\pm (L-1),...,\pm 1,0,
\end{equation}
and the weight vector components are initiated at random from a flat
distribution with equal probability for each value. These two scenarios
are referred to as the continuous case and the discrete case. 

We studied the mutual and dynamic learning of such TPMs in various
scenarios where the initial random selected weight vector is the unknown
secret information. Two machines \( A \) and \( B \), perform mutual
learning and try to synchronize by updating their weights according
to the match between their output such that at the end they achieve
full synchronization. The third machine, \( C \), performs dynamic
learning by trying to learn the weight vectors of one of the two machines,
say \( A \), and uses an attack strategy to update its weight vectors
such that at the end of the procedure they will be identical to the
weight vector of player \( A \). The application of these procedures
to the field of Cryptography is discussed in the following section.

\subsection{\label{generalFW}Cryptography Based on Synchronization: General
Framework}

Before we develop the detailed equations for mutual learning in TPMs,
we introduce the general concept of synchronization and learning in
discrete parity machines in terms of a mean-field-like approach, and
discuss the qualitative ability to construct an ephemeral key-exchange
protocol based on mutual learning between TPMs. 

First, let us consider two parties A and B who wish to agree on a
secret key over a public channel. The weight vectors, \( \W _{i}^{A/B} \),
are the parameters of each unit which are changed during the training
procedure. Both parties start with secret initial parameters \( \W  \)
which may be generated randomly. After a number of training steps,
the set of parameters is synchronized and becomes the \emph{time-dependent}
common key. At each training step a common random input \( \x _{i} \)
is generated for both of the parties; it is public and known to possible
eavesdroppers.

Each party of the secure channel consists of three hidden units with
corresponding three parameter vectors. For a given input \( \x _{i} \)
each party calculates an output bit \( \sigma ^{A/B} \) and sends
it over the public channel. A training step is performed only if the
two output bits disagree and only for the hidden units which agree
with their output\begin{equation}
\label{generalIncr}
\Delta \W ^{A/B}=g\left( \sigma ^{A/B}\x _{i}\right) \theta \left( -\sigma ^{A}\sigma ^{B}\right) \theta \left( \sigma ^{A/B}\tau ^{A/B}_{i}\right) ,
\end{equation}
 where \( g \) is an odd function. As an example consider the following
configuration of the hidden units: \( +++ \) for TPM \( A \) and
\( -++ \) for TPM \( B \). The output bits have the values \( \sigma ^{A}=1,\sigma ^{B}=-1 \).
Hence A trains all of its units according to \( \x _{i} \), while
B changes only the weight vector of its first unit according to \( -\x _{i} \).

Synchronization between the two machines indicates a full anti-parallel
state where each machine produces exactly the opposite output of the
other for any given input. The success of synchronization can be measured
by the probability of an incoherent state, i.e., the probability of
having the same output instead of the opposite one. The probability
for an \emph{incoherent state}, \( \epsilon ^{in} \), that two corresponding
hidden units are mistaken and instead of producing exactly the opposite
output they agree on a random input, is given by\begin{equation}
\label{epsgener}
\epsilon ^{in}=Prob\left( \tau ^{A}_{i}\left( \x _{i},\W ^{A}_{i}\right) =\tau ^{B}_{i}\left( \x _{i},\W ^{B}_{i}\right) \right) .
\end{equation}
The function \( g \) used for training must be chosen so that on
the average (over random input) \( \epsilon ^{in} \) is decreased.
In this section we simplify the presentation by assuming symmetry
among the three hidden unit, \( \epsilon ^{in}_{i}=\epsilon ^{in} \).
The full detailed description of the dynamical process beyond this
mean-field-like framework is given in \ref{DM}.

It is now easy to see that as soon as the TPMs are synchronized they
will remain synchronized, i.e., if \( \W _{i}^{A}=-\W _{i}^{B} \)
for all \( i, \) then \( \sigma ^{A}=-\sigma ^{B} \) and will remain
so. A training step in a unit \( i \) is performed only if both output
bits disagree and if the two \( \tau _{i} \) disagree accordingly.
Hence, after the synchronization state is achieved they either perform
a coherent training step or they do not change their parameters (referred
to as a quiet step). A pair of synchronized hidden units performs
a kind of random walk in parameter space but remains synchronized.

This is different when the two hidden units are not identical. Let
us consider the \emph{first} hidden unit, where there are four distinct
cases: 

(a) \( \sigma ^{A}=\sigma ^{B} \): nothing moves and the next step
is performed. 

(b) \( \tau ^{A}_{1}=\sigma ^{A}, \) \( \tau _{1}^{B}=\sigma ^{B} \),
\( \sigma ^{A}=-\sigma ^{B} \): both parameter vectors \( \W ^{A}_{1} \)
and \( \W ^{B}_{1} \) are coherently changed. 

(c) \( \tau ^{A}_{1}=\sigma ^{A}, \) \( \tau _{1}^{B}\neq \sigma ^{B} \),
\( \sigma ^{A}=-\sigma ^{B} \) or \( \tau ^{A}_{1}\neq \sigma ^{A}, \)
\( \tau _{1}^{B}=\sigma ^{B} \), \( \sigma ^{A}=-\sigma ^{B} \):
only one parameter vector is changed and moves incoherently, hence
\( \epsilon ^{in}_{1} \) increases.

(d) \( \tau ^{A}_{1}\neq \sigma ^{A}, \) \( \tau _{1}^{B}\neq \sigma ^{B} \),
\( \sigma ^{A}=-\sigma ^{B} \): both parameter vectors are not changed. 

The probability of finding these four cases can be calculated from
the knowledge of \( \epsilon ^{in} \). For example, the probability
of finding the configuration shown above, \( +++ \) and \( -++ \),
is \( \frac{1}{8}\left( 1-\epsilon ^{in}\right) \left( \epsilon ^{in}\right) ^{2} \).
All \( 64 \) configurations can be divided into three categories:
the probability of having an attractive step, \( p_{a} \) (case (b));
the probability of having a repulsive step, \( p_{r} \) (case (c));
or the probability of having a quiet step, \( p_{q} \) (cases (a)
and (d)). These probabilities are found to be 

\begin{eqnarray}
p_{a}=\frac{1}{2}\left[ \left( 1-\epsilon ^{in}\right) ^{3}+\left( 1-\epsilon ^{in}\right) \left( \epsilon ^{in}\right) ^{2}\right] , &  & \label{paprPM} \\
p_{r}=2\left( 1-\epsilon ^{in}\right) \left( \epsilon ^{in}\right) ^{2},\; p_{q}=1-p_{a}-p_{r}. & \nonumber 
\end{eqnarray}

In the remainder of this section the three probabilities above are
employed in order to explain the synchronization phenomenon, and to
demonstrate the superiority of the synchronization process over a
possible attacker that also tries to synchronize with \( A \) and
\( B \). 

Close to synchronization, \( \epsilon ^{in}\sim 0 \), the probability
of having a repulsive step is proportional to \( p_{r}\sim \left( \epsilon ^{in}\right) ^{2} \)
whereas the probability of having an attractive step is \( p_{a}\sim \frac{1}{2} \)
(quiet steps are always possible). Let us assume that the change of
the error, \( \epsilon ^{in} \) depends only on a function of \( \epsilon ^{in} \)
itself. Later we will derive the exact equations, which are more complex.
Then, the average change in \( \epsilon ^{in} \) in one step is obtained
by \begin{equation}
\label{deltaepsilom}
\Delta \epsilon =a\left( \epsilon ^{in}\right) p_{a}-r\left( \epsilon ^{in}\right) p_{r}.
\end{equation}
Close to synchronization a repulsive step affects all of the parameters
while an attractive step can only synchronize the few parameters which
are not yet identical. Hence we expect for small values of \( \epsilon ^{in} \):\begin{equation}
\label{tryeq}
a\left( \epsilon ^{in}\right) \sim a_{0}\epsilon ^{in},\; \; r\left( \epsilon ^{in}\right) \sim r_{0}.
\end{equation}
Therefore, in the leading order one obtains \( \Delta \epsilon \propto a_{0}\epsilon ^{in} \).
Close to synchronization the attractive force is dominate, independent
of the detailed mechanism of learning. The parity machine suppresses
the repulsive steps by reducing their appearance frequency.

This relation does not hold for the committee machine which maps the
hidden units to their majority vote, \( \sigma =sign\left( \tau _{1}+\tau _{2}+\tau _{3}\right)  \)
\cite{cm1,cm2}. For this case one finds\begin{eqnarray}
p_{a}=\frac{3}{4}\left( 1-\epsilon ^{in}\right) ^{3}+\left( 1-\epsilon ^{in}\right) ^{2}\left( \epsilon ^{in}\right) +\frac{1}{2}\left( 1-\epsilon ^{in}\right) \left( \epsilon ^{in}\right) ^{2}, &  & \\
p_{r}=\frac{1}{2}\left( 1-\epsilon ^{in}\right) ^{2}\left( \epsilon ^{in}\right) +\left( 1-\epsilon ^{in}\right) \left( \epsilon ^{in}\right) ^{2}. & \nonumber 
\end{eqnarray}
Now, close to synchronization \( p_{r}\sim \epsilon ^{in} \) and
repulsion and attractive forces are of the same order, Eq. (\ref{deltaepsilom}).
This competition between attraction and repulsion supports possible
attackers, as discussed below.

Let us go back to the parity output and consider an attacker \( C \)
who knows all the details of the algorithm and can listen to the communication
between \( A \) and \( B \). We know that the initial configurations
of the parameters of \( A \) and \( B \) are unknown. The attacker
\( C \) has the same architecture (TPM), the same number of hidden
units (\( 3 \)) and uses the same learning algorithm, Eq. (\ref{generalIncr}).
What is a good algorithm for \( C \) to synchronize, i.e., to learn
\( A \) and to be anti-parallel to \( B \)? If \( C \) is synchronized
then she should remain so. Hence she should use the identical training
step in case of agreement with \( A \). Let us consider an attacker
\( C \) who simulates party \( A \) after synchronization between
\( A \) and \( B \) is achieved. \( C \) uses the complete algorithm
explained above for party \( A \). This means that \( A \) always
makes some moves of her parameters while \( C \) moves her parameters
corresponding to the units whose output bit \( \tau _{i}^{C} \) are
identical to \( \sigma ^{A} \) (in the following we named this attack
\emph{the naive attack} - see \ref{simpleA}). This strategy for \( C \)
generates many repulsion steps between \( C \) and \( A \). In fact,
assuming the error between all matching units is the same, \( \epsilon ^{in}=Prob\left( \tau _{i}^{C}\neq \tau _{i}^{A}\right)  \)
(where we use the same symbol for \( \epsilon ^{in} \) as in Eq.
(\ref{epsgener}), although seemingly different, in both cases it
refers to the error, see \ref{OrderP} and Eq. (\ref{epsgedef}) below)
and summing up all possibilities yields\begin{eqnarray}
p_{a}=\frac{1}{2}\left( 1-\epsilon ^{in}\right) ^{3}+\frac{1}{2}\left( 1-\epsilon ^{in}\right) \left( \epsilon ^{in}\right) ^{2}+\left( 1-\epsilon ^{in}\right) ^{2}\epsilon ^{in}, &  & \label{papratt} \\
p_{r}=\left( 1-\epsilon ^{in}\right) ^{2}\epsilon ^{in}+2\left( 1-\epsilon ^{in}\right) \left( \epsilon ^{in}\right) ^{2}+\left( \epsilon ^{in}\right) ^{3}. & \nonumber 
\end{eqnarray}
The essential difference between party \( A \) and attacker \( C \)
is that the probability of finding a repulsive step scales with \( \left( \epsilon ^{in}\right) ^{2} \)
in the mutual learning between \( A \) and \( B \) and scales with
\( \epsilon ^{in} \) in the dynamic learning between \( C \) and
\( A \), close to synchronization. \( A \) and \( B \) react to
their mutual output while \( C \) cannot influence \( A \); this
yields a different behavior for small values of the error \( \epsilon ^{in} \). 

\begin{figure}
{\centering \resizebox*{0.5\textwidth}{0.3\textheight}{\rotatebox{270}{\includegraphics{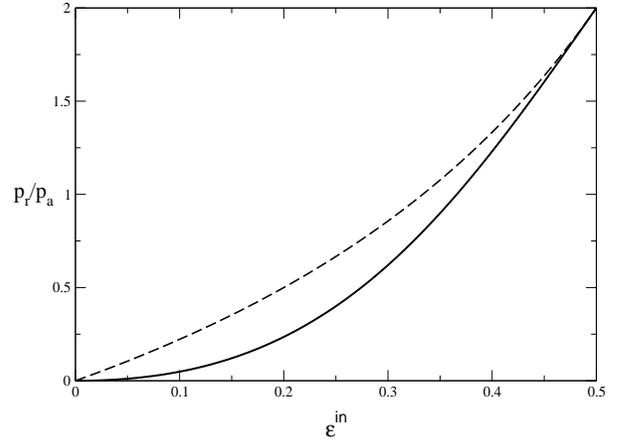}}} \par}

\caption{\label{figratio}The ratio between \protect\( p_{r}\protect \) and
\protect\( p_{a}\protect \) as a function of \protect\( \epsilon ^{in}\protect \)
in the case of mutual learning in TPMs, Eq. (\ref{paprPM}) (solid
line) and in case of the naive attack, Eq. (\ref{papratt}) (dashed
line).}
\end{figure}

The full scheme of the ratio, \( p_{r}/p_{a} \), derived from Eqs.
(\ref{paprPM}) and (\ref{papratt}) as a function of \( \epsilon ^{in} \)
is presented in Figure \ref{figratio}. It is clear that at any value
of \( \epsilon ^{in} \) the performance of the mutual learning is
better than the performance of the naive attacker that performs many
more repulsive moves compared to hers attractive moves. Therefore,
a more sophisticated attacker was recently suggested in \cite{Shamir}
- the flipping attacker. Hers performance cannot be measured in the
scope of this general framework since hers strategy depends on the
local fields in the hidden units and therefor can not be included
under the rubric of Eq. (\ref{generalIncr}), where \( g \) depends
only on \( \sigma \x _{i} \) .

In the following, before delving into details we introduce the dynamic
(Eq. (\ref{generalIncr})) more specifically. We discuss some of the
relevant order parameters and their distributions. We present the
strategy of the flipping attacker and an intuitive explanation for
her success.

\subsection{The Dynamics\label{dynamic}}

In principle, one can consider the following classes of dynamics that
lead to a synchronized state:

(A) The parties update their weight vectors whenever their outputs
mismatch (\( \sigma ^{A}\neq \sigma ^{B} \), as appears in Eq. (\ref{generalIncr})),
and each unit updates according to the input multiplied by the opposite
of its output.

(B) The parties update their weight vectors whenever their outputs
mismatch (\( \sigma ^{A}\neq \sigma ^{B} \), as appears in Eq. (\ref{generalIncr})),
and each unit updates according to the input multiplied by its output.

(C) The parties update their weight vectors whenever their outputs
match (\( \sigma ^{A}=\sigma ^{B} \)), and each unit updates according
to the input multiplied by the opposite of its output.

(D) The parties update their weight vectors whenever their outputs
match (\( \sigma ^{A}=\sigma ^{B} \)), and each unit updates according
to the input multiplied by its output.

In all the dynamics mentioned above, the \( i \)th hidden unit is
updated only if it matches the overall output in that party, if \( \tau _{i}=\sigma  \).
The two parties that try to synchronize might end up in an anti-parallel
state (cases (A) and (B)), or in a parallel state (cases (C) and (D)
). Although Eq. (\ref{generalIncr}) does not describe cases (C) and
(D), the discussion in section \ref{generalFW} is relevant to all
cases. 

In this Paper we introduce a detailed presentation of case (A). In
each step an update is made only if both machines, \( A \) and \( B \),
disagree, \( \sigma _{A}\neq \sigma _{B} \) , and each unit updates
according to the input multiplied by the opposite of its output. In
the spherical case we normalize the weight vector after each updating
such that its norm does not change. The dependence of the weight vector
in a new step on the weight\(  \) vector in the former one in the
continuous case is \begin{eqnarray}
\J _{i}^{A+}=\frac{\J _{i}^{A}+\frac{\eta }{N}\x _{i}\theta (-\sigma ^{A}\sigma ^{B})\theta (\sigma ^{A}\tau _{i}^{B})\sigma ^{B}}{\left\Vert \J _{i}^{A}+\frac{\eta }{N}\x _{i}\theta (-\sigma ^{A}\sigma ^{B})\theta (\sigma ^{A}\tau _{i}^{B})\sigma ^{B}\right\Vert }, & \label{parityEq} \\
\J _{i}^{B+}=\frac{\J _{i}^{B}+\frac{\eta }{N}\x _{i}\theta (-\sigma ^{A}\sigma ^{B})\theta (\sigma ^{B}\tau _{i}^{B})\sigma ^{A}}{\left\Vert \J _{i}^{B}+\frac{\eta }{N}\x _{i}\theta (-\sigma ^{A}\sigma ^{B})\theta (\sigma ^{B}\tau _{i}^{B})\sigma ^{A}\right\Vert }, & \nonumber 
\end{eqnarray}
 where \( \theta (y) \) is the Heavyside function, i.e., equals zero
for \( y<0 \) and \( 1 \) otherwise, \( \eta  \) is the learning
rate and \( i=1,...,K \). The analysis of the dynamic is in the thermodynamic
limit where \( N\rightarrow \infty  \) and the weight vectors are
updated by an infinitely small  quantity in each step. 

In the discrete scenario, the update is made in a similar manner,
yet there are two important differences from the dynamics point of
view. One is that in each step the vectors' components are changed
to the next integer value and not by an infinitesimally small one
as in the continuous case (Eq. (\ref{parityEq})). The second difference
is that when there is an update, the components which have reached
the boundary value \( W_{i}=\pm L \) , and their absolute value should
be increased \( W_{i}^{+}=\pm (L+1) \), are not changed, and remain
with the boundary value. Mathematically, the learning is phrased as
follows \begin{eqnarray}
\W _{i}^{A+}=\W _{i}^{A}+D(\W _{i}^{A}\x _{i}\sigma ^{B})\x _{i}\sigma ^{A}\theta (\sigma ^{A}\tau _{i}^{A})\theta (-\sigma ^{A}\sigma ^{B}), &  & \label{Wdisc} \\
\W _{i}^{B+}=\W _{i}^{B}+D(\W _{i}^{B}\x _{i}\sigma ^{A})\x _{i}\sigma ^{A}\theta (\sigma ^{B}\tau _{i}^{B})\theta (-\sigma ^{A}\sigma ^{B}), & \nonumber 
\end{eqnarray}
 where \( D(y)=1-\delta _{L,y} \) and \( \delta  \) is the Kronecker
delta function.

\subsection{Order Parameters and Joint Probability Distributions\label{OrderP}}

The analysis of learning in neural networks with an infinite number
of weight vector components is based upon statistical mechanics analysis
of several order parameters. The standard order parameters used are
\begin{eqnarray}
Q^{m}_{i}=\frac{1}{N/3}\J ^{m}_{i}\cdot \J ^{m}_{i}, &  & \\
R^{m,n}_{i}=\frac{1}{N/3}\J ^{m}_{i}\cdot \J ^{n}_{i}, & \nonumber 
\end{eqnarray}
 where the index \( i \) represents the \( i \)th hidden unit, \( i=1,...,K \)
and \( m,n \) denote the specific party, \( m,n\in \left\{ A,B,C\right\}  \).
The angle between each pair of weight vectors \( \theta  \), is given
by the normalized overlap between the weight vectors \begin{equation}
\label{setR}
\rho _{i}^{m,n}=\cos {\theta _{i}^{m,n}}=\frac{\J ^{m}_{i}\cdot \J _{i}^{n}}{\left\Vert \J ^{m}_{i}\right\Vert \left\Vert \J _{i}^{n}\right\Vert }.
\end{equation}
We assume that there are no direct correlations between different
hidden units due to the tree architecture and therefore the overlaps
between different units is zero. 

In the framework of statistical mechanics analysis of on-line learning
the order parameters play an important role in taking the averages
over the random inputs, or equivalently over the local field distribution.
According to the central limit theorem, the joint probability distribution
of the local fields in each triplet of matching hidden units taken
from the three different machines depends only on the set of order
parameters, \( P(h^{A},h^{B},h^{C}|\left\{ R,Q\right\} ) \) (where
we omitted the subscript \( i \) from all parameters) and can be
found from the correlation matrix. When all weight vectors are normalized,
\( Q^{m}=1 \), it is found to be \begin{equation}
\label{joint}
P=\frac{\exp (-\frac{F}{2E})}{(2\pi )^{3/2}\sqrt{E}},
\end{equation}
 where \( F=\left( h^{C}\right) ^{2}G^{C}+\left( h^{A}\right) ^{2}G^{A}+\left( h^{B}\right) ^{2}G^{B}-2h^{A}h^{B}D^{C}-2h^{A}h^{C}D^{B}-2h^{C}h^{B}D^{A} \),
\( E=1-\left( \rho ^{A,B}\right) ^{2}-\left( \rho ^{A,C}\right) ^{2}-\left( \rho ^{B,C}\right) ^{2}+2\rho ^{A,B}\rho ^{A,C}\rho ^{B,C} \),
\( G^{k}=(1-\rho ^{l,m})^{2} \), \( D^{k}=\rho ^{l,m}-\rho ^{k,m}\rho ^{k,l} \)
and \( k,l,m\in \left\{ A,B,C\right\}  \). This complicated expression
can be much simplified if we assume that the two machines, \( A \)
and \( B \), are already anti-parallel. In that case the joint probability
distribution of the local fields is given by \begin{equation}
\label{jointsimple}
P=\frac{e^{-\frac{1}{2}\frac{\left( h^{C}\right) ^{2}+\left( h^{A}\right) ^{2}-2h^{A}h^{C}\rho ^{A,C}}{1-\left( \rho ^{A,C}\right) ^{2}}}}{2\pi \sqrt{1-\rho ^{A,C}}}\delta (h^{A}+h^{B}),
\end{equation}
 where \( \delta () \) stand for the Dirac delta function.

At this stage it is possible to calculate the probabilities defined
in section \ref{generalFW} and to show that indeed \( \epsilon ^{in} \)
has the same meaning and the same dependency on \( \rho  \) in the
two cases: Eq. (\ref{epsgener}) and later when the attacker is introduced.
Averaging over the local field distributions results in the case of
mutual learning in \( \epsilon ^{in}=1-\frac{1}{\pi }\cos ^{-1}\rho ^{A,B} \)
and in the case of dynamic learning we find \( \epsilon ^{in}=\frac{1}{\pi }\cos ^{-1}\rho ^{A,C} \).
In order to compare these two errors, where in the first one learning
is described by negative \( \rho  \) and in the second by positive,
we define \( \bar{\rho }=|\rho ^{A,B}|=|\rho ^{A,C}| \). Substituting
\( \bar{\rho } \) into both functions above, we get\begin{equation}
\label{epsgedef}
\epsilon ^{in}=\frac{1}{\pi }\cos ^{-1}\bar{\rho }.
\end{equation}

We present in this Paper a flipping attacker, which makes use of the
absolute value of the local field. The attacker estimates that the
unit with the smallest absolute local field is the one that is most
probably wrong - that has different outputs, \( \tau _{i}^{C}\neq \tau _{i}^{A} \).
The origin of this assumption can be easily explained by averaging
over the local field distribution. The average of the absolute value
of the local field, \( \left\langle \left| h^{C}\right| \right\rangle  \),
given an overlap \( \rho ^{A,C} \) between two matching hidden units
and norm \( Q^{C} \) of the weight vector in this unit is found to
be

\begin{equation}
\label{average}
\left\langle \left| h^{C}\right| \right\rangle =\frac{1}{2}\sqrt{\frac{Q^{C}}{2\pi }}\left( 1\pm \rho ^{A,C}\right) ,
\end{equation}
where the sign in the right hand-side of the equation is plus for
agreement between the outputs and minus for disagreement. Since \( \rho  \)
varies between \( -1 \) and \( 1 \) and in a state of partial learning
\( 0<\rho <1 \), a small absolute local field signals a mistake in
the unit's output. The flipping attacker uses this knowledge in her
learning procedure, as discussed in section \ref{Geometric}. 

The analytical study of this attacker includes averages over probability
distribution of the local field in the third party, the attacker \( C \),
given the local fields of the two machines. This probability is given
by \begin{equation}
\label{dist2}
P(h^{C}|h^{B},h^{A},\left\{ \rho ,Q\right\} )=\frac{P(h^{C},h^{B},h^{C}|\left\{ \rho ,Q\right\} )}{P(h^{A},h^{B}|\left\{ \rho ,Q\right\} )}
\end{equation}
 where \( P(h^{C},h^{B},h^{C}|\left\{ \rho ,Q\right\} ) \) and \( P(h^{C},h^{B}|\left\{ \rho ,Q\right\} ) \)
are the joint probability distributions of the three local fields
and two local fields respectively, and they are derived from the correlation
matrix similar to Eq. (\ref{joint}).

In the discrete case, when the increment is finite (see for instance
Eq. (\ref{Wdisc})), the above order parameters no longer suffice
for the macroscopical description of the dynamics even in the thermodynamic
limit, \( N\rightarrow \infty  \). However, the distributions above
do hold. The dynamic cannot be analyzed with the standard equations
of motion based on differential equations of the order parameters
with respect to \( \alpha  \), the number of examples per input dimension.
We introduce a generic method for analyzing the discrete case by extending
the macroscopical parameters and deriving macro-dynamical updating
equations (see section \ref{DM}).

\section{Mapping procedure\label{mapping}}

One can map mutual learning in the parity case onto mutual learning
in \( K \) perceptrons. The mapping to noisy perceptron introduced
for analyzing on-line learning in TPM \cite{CopelliKC} is inadequate
in the case of \emph{mutual} learning where the updating depends on
the matching between the outputs but is independent of their specific
sign. Nevertheless, a different mapping from TPM to noisy perceptrons
can be used for the mutual learning case. The mapping presentation
is much simplified in the continuous case since assuming random initial
conditions to all hidden units results in the same overlap for all
hidden units, \( \rho _{i}=\rho  \) \( \forall i \). Therefore,
we first assume that all the overlaps between matching hidden units
are the same. Hence, updating \( K \) perceptrons is equivalent to
one updating in the TPM. The presentation of the mapping below is
simplified by the restriction of \( K=3 \) and the generalization
to any \( K \) is straightforward. 

We have TPMs that consist of non-overlapping receptive fields with
random inputs. Hence in each of the TPMs all \( 8 \) internal representations
appear with equal probability. A specific hidden unit is updated when
the following two conditions are fulfilled; (a) there is a mismatch
between the results of the two TPMs, and (b) the state of the hidden
unit is the same as the output of the TPM. We make use of \( \epsilon  \),
the probability of having different results in the two hidden units
that the overlap between them is \( \rho  \) and is given by\begin{equation}
\label{epsilon}
\epsilon =\frac{1}{\pi }\cos ^{-1}\rho .
\end{equation}
We concentrate on a specific pair of matched hidden units. Given that
the outputs of the hidden units are different, there is a probability,
\( P_{1} \), that the TPMs results are different and in one \textit{half}
of the cases the TPM output has the same output as its hidden unit
and therefore both hidden units in both machines are updated. This
probability is given by \begin{equation}
\label{pA}
P_{1}=P(\sigma ^{A}\neq \sigma ^{B}|\tau _{i}^{A}\neq \tau _{i}^{B})=\epsilon ^{2}+(1-\epsilon )^{2}.
\end{equation}
 Similarly, the probability that there is a mismatch between the two
TPMs given that there is agreement between two hidden units, is given
by \begin{equation}
\label{pB}
P_{2}=P(\sigma ^{A}\neq \sigma ^{B}|\tau _{i}^{A}=\tau _{i}^{B})=2\epsilon (1-\epsilon ).
\end{equation}
 In this case only one of the hidden units has the same sign as the
output in its TPM and only that hidden unit is updated.

These probabilities are introduced into the updating procedure of
the hidden units - the perceptrons. In the continuous case they affect
the form of the equations of motion (see Eq. (\ref{percept})). In
the discrete case they are introduced in a different manner, as described
in section \ref{DM}.

\section{Continuous Tree Parity Machines\label{CPM}}

Counting on the mapping procedure described above, mutual and dynamic
learning in continuous TPMs can be mapped onto learning scenarios
in continuous perceptrons. The updating rule can be redefined so that
it will be suitable for a perceptron where the kind of updating depends
on the above probabilities, \( P_{1} \) and \( P_{2} \), Eqs. (\ref{pA})
and (\ref{pB}). The standard on-line equations consist of an average
over the order parameters \cite{EnvB}, and now contain additional
random variables. The average over these additional variables is taken
by introducing auxiliary random parameters, as described in the following
section.

\subsection{Anti-Parallel Learning\label{antipa}}

In this scenario the updating rules of the TPMs are given in Eqs.
(\ref{parityEq}) where we have three hidden units, \( K=3 \). Mapping
the rules onto a perceptron learning by employing the probabilities
above is done by introducing auxiliary random parameters, \( p_{\alpha } \),
\( p_{\beta } \), \( p_{\gamma } \), which are equally distributed
between \( 0 \) and \( 1 \). The updating rule is calculated as
a function of these parameters in the following manner, \begin{equation}
\label{percept}
\W ^{A+}=\frac{\J ^{A}+\frac{\eta }{N}\x \tau ^{B}\Delta _{A}}{|\J ^{A}+\frac{\eta }{N}\x \tau ^{B}\Delta _{A}|},\; \W ^{B+}=\frac{\J ^{B}+\frac{\eta }{N}\x \tau ^{A}\Delta _{B}}{|\J ^{B}+\frac{\eta }{N}\x \tau ^{A}\Delta _{B}|},
\end{equation}
 where \begin{eqnarray}
\Delta _{A}=\theta (-\tau ^{A}\tau ^{B})\theta (\frac{P_{1}}{2}-p_{\alpha })-\theta (\tau ^{A}\tau ^{B})\theta (P_{2}-p_{\beta })\theta (\frac{1}{2}-p_{\gamma }) & , & \nonumber \label{add} \\
\Delta _{B}=\theta (-\tau ^{A}\tau ^{B})\theta (\frac{P_{1}}{2}-p_{\alpha })-\theta (\tau ^{A}\tau ^{B})\theta (P_{2}-p_{\beta })\theta (p_{\gamma }-\frac{1}{2}) & .\nonumber 
\end{eqnarray}
The introduction of the auxiliary random variables is done according
to the following logic: in one half of the cases of disagreement between
the units and disagreement between the TPMs, no update occurs in the
units (since their sign does not match the TPM's sign) and hence \( P_{1} \)
is divided by \( 2 \) in the equation above. The second scenario
where updating occurs is when the units have the same sign, the TPMs
disagree and therefore one of the units is updated and the other is
not. The auxiliary random number \( p_{\gamma } \) is the one that
determines (randomly) which unit of the two is updated.

\begin{figure}
{\centering \resizebox*{0.5\textwidth}{0.3\textheight}{\rotatebox{270}{\includegraphics{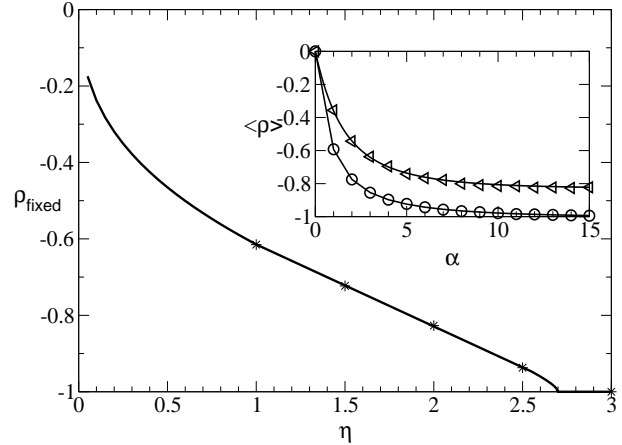}}} \par}

\caption{\label{eta}The fixed point \protect\( \rho _{f}\protect \) as a
function of \protect\( \eta \protect \) for the continuous TPM as
obtained from the solution of Eq. (\ref{fix}) (solid line). Simulation
results in some instances of \protect\( \eta \protect \) are presented
by stars. Inset: Analytical (solid lines) and simulation results in
the case of \protect\( \eta =2\protect \) (triangles) and \protect\( \eta =3\protect \)
(circles) for \protect\( \left\langle \rho \right\rangle \protect \)
as a function of \protect\( \alpha \protect \). All simulations are
carried out with \protect\( N=5000\protect \) and averaged over \protect\( 20\protect \)
samples. }
\end{figure}

In order to calculate the equations of motion, one has to multiply
the updated vectors, Eq. (\ref{percept}), first, and then to perform
the two averages; average over the joint probability distributions
of the local fields and over the random parameters, \( p_{\alpha } \),
\( p_{\beta } \) and \( p_{\gamma } \). The result of these two
averages is an equation over the normalized overlap \( \rho  \),
that depends only on \( \rho  \) or equivalently on the angle, \( \theta  \),
(see Eq. (\ref{setR})) \begin{eqnarray}
\frac{d\rho }{d\alpha }=\eta [\frac{\theta ^{2}}{\pi ^{2}}+(1-\frac{\theta }{\pi })^{2}][\frac{1}{\sqrt{2\pi }}(1-\rho )-\frac{\eta \theta }{2\pi }](1+\rho ) &  & \label{rho} \\
-\frac{2\eta }{\sqrt{2\pi }}(1-\rho ^{2})\frac{\theta }{\pi }(1-\frac{\theta }{\pi })-\eta ^{2}\rho \frac{\theta }{\pi }(1-\frac{\theta }{\pi })^{2}, & \nonumber 
\end{eqnarray}
where \( \alpha  \) is the number of examples per input dimension.
The points \( \rho =\pm 1 \) are fixed points of the equation of
motion above. Both are repulsive when the learning rate, \( \eta  \),
is small. As soon as \( \eta >\eta _{c}\sim 2.68 \) a phase transition
occurs, the \( \rho =-1 \) fixed point becomes an attractive one
and a new phase arises, where the two machines are fully synchronized.
The asymptotic decay of \( \rho  \) to synchronization scales exponentially
with \( \alpha  \), as can be found by expanding the terms in Eq.
(\ref{rho}) around \( \theta =\pi  \). Apart from the fixed points
discussed above, for any \( \eta  \) smaller than \( \eta _{c} \)
there is a different attractive fixed point, as can be found by solving
numerically Eq. (\ref{rho}). The fixed point \( \theta _{f}(\rho _{f}) \)
is the exact angle(overlap) in a specific learning rate, \( \eta  \),
in which the right hand side of equation \ref{rho} becomes zero:

\begin{equation}
\label{fix}
\eta =\frac{\frac{\sqrt{2\pi }}{\theta _{f}}\sin {^{2}\theta _{f}}(1-\frac{2\theta _{f}}{\pi })^{2}}{(1+\cos {\theta _{f}})(\frac{\theta _{f}^{2}}{\pi ^{2}}+(1-\frac{\theta _{f}}{\pi })^{2})+2\cos {\theta _{f}}(1-\frac{\theta _{f}}{\pi })^{2}}.
\end{equation}

In Figure \ref{eta} we plotted the fixed points as a function of
\( \eta  \), as was found numerically from Eq. (\ref{fix}). Simulation
results for spherical TPMs with \( N=5000 \) and averaged over \( 20 \)
samples are in agreement with the analysis as indicated by the few
tested cases presented by the symbols. Clearly, the system undergoes
a phase transition from a partial to a perfect anti-parallel state
at \( \eta _{c}\sim 2.68 \). One instance for each of the phases
is given in the inset of Figure \ref{eta}. The development of the
averaged \( \left\langle \rho \right\rangle  \), averaged over the
three hidden units and 20 samples, in the case of partial mutual learning,
\( \eta =2 \) (triangles), and the case of anti-parallel synchronization,
\( \eta =3 \) (circles), as a function of \( \alpha  \) is presented
in the inset of Figure \ref{eta}. Numerical calculations of the analytical
equation, Eq. (\ref{rho}), are presented by the solid lines.

\subsection{\label{dylear}Dynamic Learning}

In the last section we show a procedure that leads to full synchronization.
In the following we check the ability of a third TPM, an attacker,
to learn the weight vectors of the two parties. The third machine,
\( C \), that tries to imitate \( A \), updates its weight vector
only when the two parties are updated and only the hidden units that
match the output of party \( A. \) Mathematically, this is defined
as follows

\begin{eqnarray}
\J _{i}^{C+}=\frac{\J _{i}^{C}+\frac{\eta }{N}\x _{i}\theta (-\sigma ^{A}\sigma ^{B})\theta (\sigma ^{A}\tau _{i}^{C})\sigma ^{B}}{\left\Vert \J _{i}^{C}+\frac{\eta }{N}\x _{i}\theta (-\sigma ^{A}\sigma ^{B})\theta (\sigma ^{A}\tau _{i}^{C})\sigma ^{B}\right\Vert } & . & \label{parity} 
\end{eqnarray}
Continuing the same line of introducing probabilities in the mutual
learning procedure, one can write a set of updating rules for the
dynamic and mutual learning in perceptrons which is equivalent to
TPMs learning. This is given by \begin{eqnarray}
\J ^{A+}=\frac{\J ^{A}+\frac{\eta }{N}\x \tau ^{B}\tilde{\Delta }_{A}}{\left\Vert \J ^{A}+\frac{\eta }{N}\x \tau ^{B}\tilde{\Delta }_{A}\right\Vert }, & \label{percop} \\
\J ^{B+}=\frac{\J ^{B}+\frac{\eta }{N}\x \tau ^{A}\tilde{\Delta }_{B}}{\left\Vert \J ^{B}+\frac{\eta }{N}\x \tau ^{A}\tilde{\Delta }_{B}\right\Vert } & , & \nonumber \\
\J ^{C+}=\frac{\J ^{C}+\frac{\eta }{N}\x \tau ^{B}\Delta _{C}}{\left\Vert \J ^{C}+\frac{\eta }{N}\x \tau ^{B}\Delta _{C}\right\Vert }, & \nonumber 
\end{eqnarray}
 where \begin{eqnarray}
\tilde{\Delta }_{A}=\theta (-\tau ^{A}\tau ^{B})\theta (P_{1}-p_{\alpha })\theta (\frac{1}{2}-p_{\delta }) &  & \nonumber \label{add2} \\
+\theta (\tau ^{A}\tau ^{B})\theta (P_{2}-p_{\beta })\theta (\frac{1}{2}-p_{\gamma }), &  & \nonumber \\
\tilde{\Delta }_{B}=\theta (-\tau ^{A}\tau ^{B})\theta (P_{1}-p_{\alpha })\theta (\frac{1}{2}-p_{\delta }) &  & \nonumber \\
+\theta (\tau ^{A}\tau ^{B})\theta (P_{2}-p_{\beta })\theta (p_{\gamma }-\frac{1}{2}), &  & \nonumber \\
\Delta _{C}=\theta (-\tau ^{A}\tau ^{B})\theta (\tau ^{A}\tau ^{C})\theta (P_{1}-p_{\alpha })\theta (\frac{1}{2}-p_{\delta }) &  & \nonumber \\
+\theta (\tau ^{A}\tau ^{B})\theta (\tau ^{A}\tau ^{C})\theta (P_{2}-p_{\beta })\theta (\frac{1}{2}-p_{\gamma }) &  & \nonumber \\
-\theta (-\tau ^{A}\tau ^{B})\theta (-\tau ^{A}\tau ^{C})\theta (P_{1}-p_{\alpha })\theta (p_{\delta }-\frac{1}{2}) &  & \nonumber \\
+\theta (\tau ^{A}\tau ^{B})\theta (-\tau ^{A}\tau ^{C})\theta (P_{2}-p_{\beta })\theta (p_{\gamma }-\frac{1}{2}). &  & \nonumber 
\end{eqnarray}
 We introduce another random parameter, \( p_{\delta } \), which
is redundant when one calculates only the mutual learning, Eq. (\ref{percept}),
and it is necessary for deriving equations of motion for the order
parameters in the case of dynamic learning. The four terms in \( \Delta _{C} \)
represent the four possibilities that cause an updating in the attacker
hidden unit. For instance, the first term of \( \Delta _{C} \) represents
the case where the hidden unit in the attacker and in the first TPM
have the same state, the TPMs' outputs are different (indicated by
\( P_{1} \)) and the outputs in the hidden units of \( A \) and
\( B \) are the same as their TPMs, (the probability for such an
event is \( \frac{1}{2} \)). 

The equation of motion after synchronization, i.e., when \( \rho _{A,B}=-1 \),
\( \rho _{A,C}=-\rho _{B,C} \), is derived by averaging Eqs. (\ref{percop})
over the joint probability distributions that is given in Eq. (\ref{jointsimple}).
It depends on the learning rate and the overlap \( \rho _{A,C} \)
and is given explicitly by \begin{equation}
\label{opprho}
\frac{d\rho _{A,C}}{d\alpha }=\frac{\eta ^{2}}{2}\left( 1-\frac{1}{\pi }\cos ^{-1}{\rho _{A,C}}-\rho _{A,C}\right) .
\end{equation}
 This equation describes the development of the overlap between the
attacker and one of the two machines that are synchronized in both
cases, when each machine learns the opposite of its result, Eq. (\ref{parity}).

\begin{figure}
{\centering \resizebox*{0.5\textwidth}{0.3\textheight}{\rotatebox{270}{\includegraphics{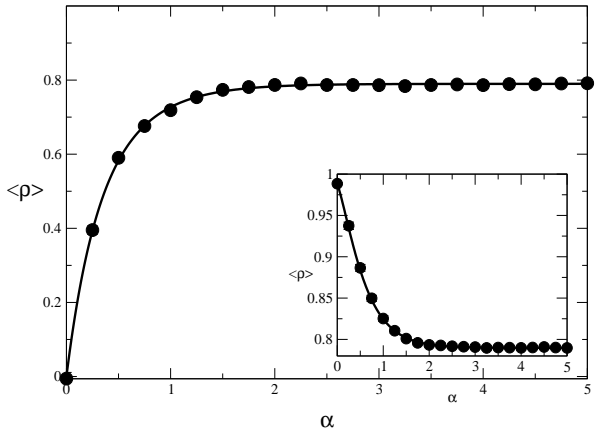}}} \par}

\caption{\label{opo}The analytical curve of the averaged overlap, \protect\( \left\langle \rho \right\rangle \protect \),
in a dynamic learning of TPMs as obtained from Eqs. (\ref{opprho})
(solid line), with \protect\( \eta =10\protect \). The initial state
is \protect\( \rho =0\protect \). Inset: Analytical results for the
dynamic learning with the initial state \protect\( \rho =0.98\protect \).
Symbols represent the corresponding simulations, carried out with
\protect\( N=5000\protect \) and averaged over \protect\( 20\protect \)
runs.}
\end{figure}

As can be derived from Eq. (\ref{opprho}), independent of the learning
rate, \( \eta , \) there is a unique fixed point \( \rho _{f}\sim 0.79 \).
The point \( \rho =1 \) is not a fixed point at all. Note that this
fixed point describes only the failure of the continuous attacker,
the equivalent \emph{discrete} attacker might synchronize and gain
\( \rho =1 \), as discussed in section \ref{simpleA}. In Figure
\ref{opo} we present analytical (solid lines) and simulation results
(symbols) for the overlap between that attacker and player A, \( \rho _{AC} \).
We carried out simulations with \( N=5000 \), and each result averaged
\( 20 \) times. A good agreement between simulation results and analytical
results is presented in Figure \ref{opo} in both cases; when the
overlap is initialized zero, \( \rho _{AC}=0 \) and in the inset,
when the initial value of the overlap is almost \( 1 \), \( \rho _{AC}=0.98 \).
All results are for full synchronization between \( A \) and \( B \),
\( \rho _{AB}=-1 \).

\subsection{Summary\label{summcont}}

In summary, we showed that an initiated pair of random TPMs that perform
mutual learning results in a full synchronization state for \( \eta >\eta _{c} \).
We introduce here a specific dynamic where the parties update only
in a mismatch between the outputs, the updating is in opposite directions
of each other and they are normalized in each step (case A in \ref{dynamic}).
Analyzing case B, for instance, reveals that for all \( \eta  \),
the stationary solution is a synchronized state. Using the dynamics
appearing in \ref{dynamic} but without normalizing the weight vectors
does not end in a synchronization state at all. The specific algorithm
we chose contains the reach phenomenon of phase transition \cite{PhaseTrans}.
Moreover, its synchronization abilities are closely related to the
discrete synchronization studied in the following section.

The attacker tries to learn the parties' weight vectors but manages
to achieve only partial success. This difficulty in learning that
such a naive attacker faces as indicated by the fixed point that differs
from \( 1 \), also characterizes the naive attacker in the other
cases presented in \ref{dynamic}. However, the analysis is not relevant
for the discrete case studied below. In the discrete case the naive
attacker performance is restricted too but perfect learning is possible,
see \ref{simpleA}. The flipping attacker that makes use of the local
fields (see \ref{Geometric}) has a better performance in the discrete
case. An open question which deserves further research, is how to
analyze the continuous flipping attacker.

\section{Discrete Machines\label{DM}}

The study of discrete networks requires different methods of analysis
than those used for the continuous case. We found that instead of
examining the evolution of \( R \) and \( Q \), we must examine
\( (2L+1)\times (2L+1) \) parameters, which describe the mutual learning
process. By writing a Markovian process that describes the development
of these parameters, one gains an insight into the learning procedure.
Thus we define a \( (2L+1)\times (2L+1) \) matrix, \( \mathbf{F}^{\mu } \),
in which the state of the machines in the time step \( \mu  \) is
represented. The elements of \( \mathbf{F} \), are \( f_{qr} \),
where \( q,r=-L,...-1,0,1,...L \). The element \( f_{qr} \) represents
the fraction of components in a weight vector in which the \( A \)'s
components are equal to \( q \) and the matching components in d
unit \( B \) are equal to \( r \). Hence, the overlap between the
two units as well as their norms are defined through this matrix,
\begin{eqnarray}
R=\sum _{q,r=-L}^{L}qrf_{qr}, &  & \label{order} \\
Q^{A}=\sum _{q=-L}^{L}q^{2}f_{qr}\; \; \; Q^{B}=\sum _{r=-L}^{L}r^{2}f_{qr}.\nonumber 
\end{eqnarray}

The matrix elements are updated, if and only if, an update of the
weight vectors occurs.

\subsection{\label{discPerc}Learning with Discrete Perceptrons}

The mutual learning scenario is much simplified in the case of the
perceptron, therefore we present here the full description of the
analytical procedure used for this case. Updating is done in the case
of a mismatch, and the aim is to arrive at a state in which the weight
vectors are anti-parallel, \( \rho =-1 \) (we could aim at \( \rho =1 \)
instead, see the manifold of possible dynamics in \ref{generalFW},
and the results would be equivalent). The dependence of the weight
vector in a new step on the weight vector in the former one is given
by:\begin{eqnarray}
\W _{i}^{A+}=\W _{i}^{A}+D(\W _{i}^{A}\x _{i}\sigma ^{B})\x _{i}\sigma ^{B}\theta (-\sigma ^{A}\sigma ^{B}), &  & \label{W} \\
\W _{i}^{B+}=\W _{i}^{B}+D(\W _{i}^{B}\x _{i}\sigma ^{A})\x _{i}\sigma ^{A}\theta (-\sigma ^{A}\sigma ^{B}), & \nonumber 
\end{eqnarray}
where \( \sigma ^{A/B} \) represents the output of TPM \( A/B \),
and \( \W ^{A/B} \) represents its weight vector.

The update of the elements of the matrix \( \mathbf{F} \), is calculated
directly from Eq. (\ref{W}), where one must average over the input
components \( X_{ij} \). On the average, half of the updated weights
in one machine are increased by 1, while the matching weights in the
other machine are decreased by 1 and vice versa. 

The possibility for agreement/disagreement between the parties is
a function of the current overlap between them, calculated using the
matrices (see Eq. (\ref{order})). This probability is implemented
by choosing a random parameter, \( p_{\alpha } \) between \( [0,1] \).
If it is smaller than \( \epsilon  \), as defined in Eq. (\ref{epsilon}),
the parties disagree, otherwise they agree. The updating of matrix
elements is described as follows: for the elements with \( q \) and
\( r \) which are not on the boundary, (\( q\neq \pm L \) and \( r\neq \pm L \))
the update can be written in a simple manner, \begin{equation}
\label{percmat}
f_{q,r}^{+}=\theta \left( p_{\alpha }-\epsilon \right) f_{q,r}+\theta \left( \epsilon -p_{\alpha }\right) \left( \frac{1}{2}f_{q+1,r-1}+\frac{1}{2}f_{q-1,r+1}\right) .
\end{equation}
 For elements with both indices on the boundary, the update is \begin{eqnarray}
f_{L,L}^{+} & = & \theta \left( p_{\alpha }-\epsilon \right) f_{L,L},\label{boundary} \\
f_{-L,-L}^{+} & = & \theta \left( p_{\alpha }-\epsilon \right) f_{-L,-L},\nonumber \\
f_{L,-L}^{+} & = & \theta \left( p_{\alpha }-\epsilon \right) \left( \frac{1}{2}f_{L,-L}\right) +\theta \left( \varepsilon -p_{\alpha }\right) \times \nonumber \\
 &  & \left( \frac{1}{2}f_{L-1,-L+1}+\frac{1}{2}f_{L-1,-L}+\frac{1}{2}f_{L,-L+1}\right) ,\nonumber \\
f_{-L,L}^{+} & = & \theta \left( p_{\alpha }-\epsilon \right) f_{-L,L}+\theta \left( \epsilon -p_{\alpha }\right) \times \nonumber \\
 &  & \left( \frac{1}{2}f_{-L+1,L-1}+\frac{1}{2}f_{-L+1,L}+\frac{1}{2}f_{-L,L-1}\right) .\nonumber 
\end{eqnarray}
 For elements with just one of the indices on the boundary (\( q=\pm L \)
and \( r\neq \pm L \) or vice versa), the update is \begin{eqnarray}
f_{q,L}^{+}=\theta \left( p_{\alpha }-\epsilon \right) f_{q,L}+ &  & \label{boundaryh} \\
\theta \left( \epsilon -p_{\alpha }\right) \left( \frac{1}{2}f_{q+1,L-1}+\frac{1}{2}f_{q+1,L}\right) , &  & \nonumber \\
f_{q,-L}^{+}=\theta \left( p_{\alpha }-\epsilon \right) f_{q,-L}+ &  & \nonumber \\
\theta \left( \epsilon -p_{\alpha }\right) \left( \frac{1}{2}f_{q-1,-L+1}+\frac{1}{2}f_{q-1,-L}\right) , &  & \nonumber \\
f_{L,r}^{+}=\theta \left( p_{\alpha }-\epsilon \right) f_{L,r}+ &  & \nonumber \\
\theta \left( \epsilon -p_{\alpha }\right) \left( \frac{1}{2}f_{L-1,r+1}+\frac{1}{2}f_{L,r+1}\right) , &  & \nonumber \\
f_{-L,r}^{+}=\theta \left( p_{\alpha }-\epsilon \right) f_{-L,r}+ &  & \nonumber \\
\theta \left( \epsilon -p_{\alpha }\right) \left( \frac{1}{2}f_{-L+1,r-1}+\frac{1}{2}f_{-L,r-1}\right) , & \nonumber 
\end{eqnarray}

The main quantity of interest is the number of steps required in order
to arrive at a state of full synchronization. In simulations there
is a discrete transition from an overlap which is almost anti-parallel
to a completely anti-parallel state. This is due to the finite nature
of the vectors, the largest value of overlap before synchronization
is \( -1+O(1/N) \). In simulations with \( N=10^{4} \), for example,
the largest value of the overlap before full synchronization is \( \rho =0.99999 \),
and this is the value we used in our analytical procedure, for defining
full synchronization for comparison to simulations with \( N=10^{4} \). 

\begin{figure}
{\centering \resizebox*{0.5\textwidth}{0.3\textheight}{\rotatebox{270}{\includegraphics{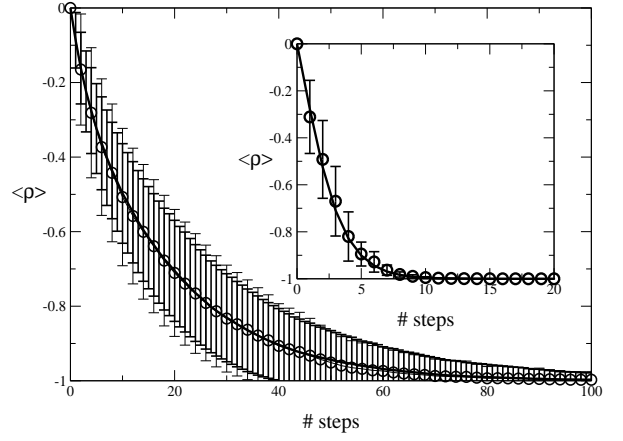}}} \par}

\caption{\label{discrete}The averaged overlap \protect\( \left\langle \rho \right\rangle \protect \)
and its standard deviation as a function of the number of steps as
found from the analytical results (solid line) and simulation results
(circles) of mutual learning in TPMs. Inset: analytical results (solid
line) and simulation results (circles) results for the perceptron,
with \protect\( L=1\protect \) and \protect\( N=10^{4}\protect \). }
\end{figure}

Our results indicate that the order parameters are not self-averaged
quantities \cite{Reens}. Several runs with the same \( N \), results
in different curves for the order parameters as a function of the
number of steps, see Figure \ref{discrete}. This explains the non-zero
variance of \( \rho  \) as a results of the fluctuations in the local
fields induced by the input even in the thermodynamic limit.

In the inset of Figure \ref{discrete} we present the averaged numerical
results derived from the analytical equations, (\ref{percmat}, \ref{boundary},
\ref{boundaryh}) of synchronization in the perceptron (solid line)
with \( L=1 \), \( W_{i}=\pm 1,0 \), . The analytical results are
averaged over \( 500 \) samples and the non-zero standard deviations
are not presented in order to simply the presentation. Simulation
results with \( L=1 \) (\( W_{i}=\pm 1,0 \)) and \( N=10^{4} \),
averaged over \( 500 \) samples are presented by the circles; error
bars are standard deviations. Note that even though the matrix elements
were initiated with the same values in each run, there is still a
non-zero standard deviation due to fluctuations in the local fields
as a function of the particular set of random inputs even in the thermodynamic
limit.

For the perceptron, synchronization is much easier and faster to achieve
than for the TPM. Take for example the case where \( L=1 \). If for
three consecutive steps, both the other party's output and \( x_{i} \)
were positive, an attacker can surely know that \( W_{i}=1 \), while
this is not so in the TPM case, as the attacker cannot know for sure
whether the unit was updated or not. Therefore, the TPM is much more
suitable for building a cryptosystem than the perceptron.

\subsection{\label{discPM}Synchronization in TPMs}

Mutual learning in discrete TPMs is described by mutual learning discrete
noisy perceptrons. As the TPM consists of three hidden units (each
evolving differently), we now have three different angles, \( \theta _{i} \)
where \( i=1,2,3 \), for each hidden unit. Since the dynamics are
not self-averaged, we use probabilities similar to those introduced
in Eq. (\ref{pA}). The definitions of these probabilities are extended
to include all three hidden units, and each one is characterized by
its own angle, \( P^{i}_{1} \), \( P^{i}_{2} \). The probability
of \( P_{1}(\sigma ^{A}\neq \sigma ^{B}|\tau _{i}^{A}\neq \tau _{i}^{B}) \),
is given by \begin{equation}
\label{pAi}
P^{i}_{1}=\epsilon _{j}\epsilon _{k}+(1-\epsilon _{j})(1-\epsilon _{k}).
\end{equation}
 Similarly, the probability that there is a mismatch between the two
TPMs given that there is agreement between the \( i \)th pair of
hidden units, for instance, is given by \begin{equation}
\label{pBi}
P^{i}_{2}=\epsilon _{j}(1-\epsilon _{k})+\epsilon _{k}(1-\epsilon _{j}).
\end{equation}
 Here, as well as in the continuous case, we chose a sequence of random
parameters to represent the particular choice of random inputs.

We follow each hidden unit separately and therefore we have three
matrices, \( \mathbf{F}^{i} \). We initialize the weights randomly,
therefore the matrices in the initial state have the values of \( 1/(2L+1)^{2} \)
in each entry. In each step, two sets of random parameters are chosen
and are used to set a specific realization of the internal presentation
for the parties. The first set is used to define agreement or disagreement
between each pair of hidden units, as done in the perceptron case
\ref{discPerc}. 

All in all, due to inversion symmetry, when \( K=3 \) there are four
possible results for the internal presentations, \( +++ \), \( +-- \),
\( -+- \) or \( --+ \) and accordingly \( 4\times 4 \) possible
states, for which the parties' output does not match, and an update
is performed. We then use the second set of random parameters for
defining the specific internal presentation in one of the TPMs, and
therefore immediately in the other, according to their agreement/disagreement.

The case when the three hidden units disagree is exemplified below.
There is a possibility that all hidden units are updated, (case (b)
in \ref{generalFW}), or only one of them; (case (b) describes two
of the hidden units and case (d) describes the third). In two of the
eight such internal presentations all the three hidden units are updated
whereas in the other six, only one of them is updated, so that we
must choose which one. All of these possibilities are equally probable,
independent of \( {\theta _{i}} \). Therefore, we take all the possible
internal scenarios into account and, for instance, when after using
the auxiliary random numbers, all three hidden units disagree, we
choose at random \( p_{\alpha } \) and accordingly update, \begin{eqnarray}
f_{q,r}^{i+} & = & \theta (\frac{1}{4}-p_{\alpha })(\frac{1}{2}f^{i}_{q+1,r-1}+\frac{1}{2}f^{i}_{q-1,r+1})\label{disc1} \\
 &  & +\theta (\frac{i+1}{4}-p_{\alpha })\theta (p_{\alpha }-\frac{i}{4})(\frac{1}{2}f^{i}_{q+1,r-1}+\frac{1}{2}f^{i}_{q-1,r+1}).\nonumber \\
 & \nonumber 
\end{eqnarray}
 The first term corresponds to the case where all three hidden units
are updated (with probability \( \frac{1}{4} \)). The second term
corresponds to the case where only one hidden unit is updated. Eq.
(\ref{disc1}) is valid only for \( q \) and \( r \) which are not
on the boundary.

\begin{figure}
{\centering \resizebox*{0.5\textwidth}{0.3\textheight}{\rotatebox{270}{\includegraphics{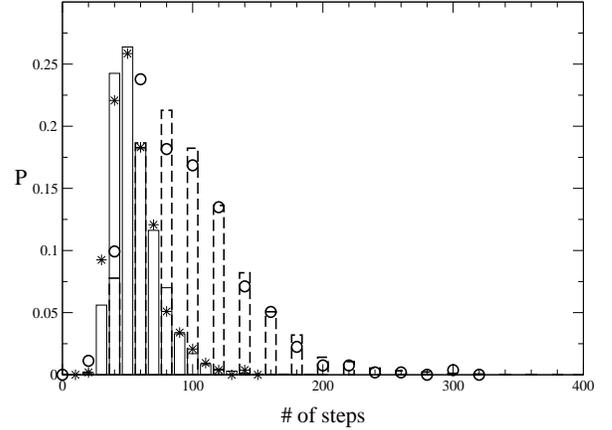}}} \par}

\caption{\label{dist1}The synchronization time (dashed line) and the dynamic
learning time (solid line) distribution, of analytical results for
TPMs, with \protect\( L=1\protect \). Symbols stand for the simulations
results, with \protect\( N=10000\protect \). }
\end{figure}

In the case of the perceptron when an update occurs, both sides perform
the update, in opposite directions. In the case of the TPMs, two matching
units do not always perform an update together; in many cases one
of the parties updates unit \( i \), while the other updates unit
\( j \), \( i\neq j \), as described in case (c) in \ref{generalFW}.
In such a case, Eq. (\ref{disc1}) is not sufficient, and we should
add a description of the matrices' update when only one party is updated.
Let us say the party represented by the matrix rows is updated. Then
we have \begin{equation}
\label{disc12}
f_{q,r}^{i+}=\frac{1}{2}f^{i}_{q+1,r}+\frac{1}{2}f^{i}_{q-1,r},
\end{equation}
and if the party represented by the matrix columns is updated, we
have \begin{equation}
\label{disc13}
f_{q,r}^{i+}=\frac{1}{2}f^{i}_{q,r+1}+\frac{1}{2}f^{i}_{q,r-1},
\end{equation}
where we limit the description only to \( q,r \) which are not on
the boundary. An example is the case when the internal presentation
of party \( A \) is \( -++ \) and that of \( B \) is \( --+ \).
Then party \( A \) updates unit \( 1 \), Eq. (\ref{disc12}) with
\( i=1 \), while party \( B \) updates unit \( 3 \), Eq (\ref{disc13})
with \( i=3 \).

In Figure \ref{dist1} we present the distribution of time steps for
synchronization according to simulations with \( N=10,000 \), (\( \star  \)),
and according to the analytical results (solid line) in the case of
\( L=1 \), taken from \( 500 \) different runs. The evolution of
the average overlap in this case is given in Figure \ref{discrete}.
A solid line represents the analytical results and circles stand for
simulation results. Both standard deviations are indicated by the
error bars. There is good agreement between the analytical and simulation
results.

An attacker does not have to achieve full synchronization in order
to decipher the secret code. For finite \( N \), even a state close
enough to synchronization is sufficient to break the code, thus making
the system insecure. Moreover, the analysis and the simulations are
faster when the aim is to arrive at a partial overlap state. We therefore
considered an attacker who achieves \( \left\langle \rho \right\rangle =0.9 \),
a successful attacker, and synchronization and learning times given
in Figure \ref{times_dist} and in Table \ref{tbl1} are for achieving
\( \left\langle \rho \right\rangle =0.9 \).

\begin{table}
\begin{tabular}{|c|c|c|c|}
\hline 
&
\( t_{synch} \)&
\( t_{naive} \)&
\( t_{flipp} \)\\
\hline
\hline 
\( L=1 \)&
\( 25\pm 14 \)&
\( 36\pm 18 \)&
\( 32\pm 19 \)\\
\hline 
\( L=2 \)&
\( 79\pm 38 \)&
\( 239\pm 145 \)&
\( 108\pm 58 \)\\
\hline 
\( L=3 \)&
\( 166\pm 67 \)&
\( 3320\pm 3039 \)&
\( 221\pm 106 \)\\
\hline 
\( L=4 \)&
\( 298\pm 113 \)&
\( 176810\pm 179,446 \)&
\( 380\pm 159 \)\\
\hline
\end{tabular}

\caption{\label{tbl1}Average synchronization and dynamic learning times,
for the naive attacker and the flipping attacker, for different values
of \protect\( L\protect \).}
\end{table}

\subsection{\label{simpleA}The Naive Attacker}

The aim of an attacker is to synchronize with one of the parties and
reveal the secret key (the weights of the parties), hence her natural
strategy is to imitate one of them, party \( A \) for instance, by
using the same learning rule. The attacker, eavesdropping on the public
channel connecting the parties, knows the input vector \( \x _{i} \)
and the output \( O^{A/B} \). When \( O^{A}\neq O^{B} \), the parties
update their weights, and so does the attacker. In the case where
the attacker's internal presentation is the same as \( A \)'s, they
update the same units, an attractive step occurs, and the attacker
gets closer to her goal. Yet when the internal presentations of the
attacker and the party differ, she updates some wrong units, a repulsive
step occurs, and this delays her. The \( 2^{K-1} \)-fold degeneracy
in the output is the main reason for the attacker's failure. The dependence
of the attacker's weight vector in a new step on the weight vector
in the former one is given by\begin{equation}
\label{discreteOP}
\W _{i}^{C+}=\W _{i}^{C}+D(\W _{i}^{C}\x _{i}\sigma ^{B})\x _{i}\sigma ^{B}\theta (-\sigma ^{A}\sigma ^{B}).
\end{equation}
The analysis is similar to the synchronization process, given by Eq.
(\ref{disc1}). We now create \( 9 \) matrices, each representing
the state of two matching hidden units among two parties, and the
attacker and each party. We must set the parties' internal presentation,
as well as the attacker's. We decide which one of the \( 8\times 8\times 8 \)
internal presentations occurs in each step, following the correlation
between the parties and the attacker, and update the matrices accordingly,
as described in \ref{discPM}. 

Although the attacker may synchronize before the parties, the average
learning time is around twice the synchronization time for \( L=1 \),
and is around \( 200 \) times the synchronization time for \( L=3 \).
It seems that the reason for the naive attacker's weakness is that
too many repulsive steps occur; therefore, when trying to improve
her abilities, we need to increase the probability for an attractive
step, and decrease the probability for a repulsive one. It has been
shown \cite{Liat&Kanter} that a small absolute local-field value
indicates a high probability for an error. In the next section we
present an advanced attacker which makes use of this knowledge.

\subsection{The Flipping Attacker\label{Geometric}}

The flipping attacker's strategy, recently introduced in \cite{Shamir},
adds a different move to the strategy of the naive attacker when disagreement
occurs between the outputs of the attacker and party \( A \). In
this case, the attacker is certain that either one or three of her
hidden units are in disagreement with \( A \)'s units, and therefore
a repulsive step will occur. Since disagreement of three units is
less likely than disagreement of one unit, the attacker's strategy
treats all cases as a one unit disagreement. The flipping attacker
tries to prevent the repulsive step by using a \char`\"{}flipping\char`\"{}
approach; she negates the sign of one of her units, before performing
the update. If the correct unit was chosen, then the \char`\"{}new\char`\"{}
internal presentation matches that of the party, and the same units
will be updated by both, thus performing an attractive step. To raise
her chances of flipping the right unit, the attacker chooses the one
whose absolute local-field value is the lowest of the three : \( \hat{\tau }_{i}=-\tau _{i} \)
for \( i \) that minimizes |\( h_{i} \)|.

The learning rules are the same as those given by Eq. (\ref{Wdisc})
for the mutual synchronization, but the attacker's learning is different,
\begin{eqnarray}
 & \label{Wnew} \\
\W _{i}^{C+} & = & \W _{i}^{C}+D(\W _{i}^{C}\x _{i}\sigma ^{B})\x _{i}\sigma ^{B}\theta (-\sigma ^{A}\sigma ^{B})\times \nonumber \\
 &  & [\theta (\sigma ^{C}\sigma ^{A})\theta (\sigma ^{C}\tau _{i}^{C})+\theta (-\sigma ^{C}\sigma ^{A})\theta (\sigma ^{A}\hat{\tau }_{i}^{C})]\nonumber \\
 & \nonumber 
\end{eqnarray}
where \( \hat{\tau }_{i}=-\tau _{i} \) if \( |h_{i}|<|h_{j}| \),
\( \forall j\neq i \) and \( \hat{\tau }_{i}=\tau _{i} \) otherwise.

The analysis used here is the same as for the naive attacker. Here
too, we follow the development of \( 9 \) matrices which are updated
at every time step, as described in \ref{discPM}. However, in cases
where the attacker's output disagrees with the A's output, we compute
the probability for every unit to be the one with the lowest absolute
local field value. For instance, when \( h^{C}_{i}>0,\forall i \)
, the probability for \( h_{1} \) being the smallest is given by:

\begin{eqnarray}
P(h^{C}_{1}<h^{C}_{2},h^{C}_{1}<h^{C}_{3})= &  & \label{probLF1} \\
\int ^{\infty }_{0}P(h^{C}_{1}|h^{A}_{1},h^{B}_{1},\left\{ \rho ,Q\right\} )dh^{C}_{1} &  & \nonumber \label{probLF} \\
\int ^{\infty }_{h^{C}_{1}}P(h^{C}_{2}|h^{A}_{2},h^{B}_{2},\left\{ \rho ,Q\right\} )dh^{C}_{2} &  & \nonumber \\
\int _{h^{C}_{1}}^{\infty }P(h^{C}_{3}|h^{A}_{3},h^{B}_{3},\left\{ \rho ,Q\right\} )dh^{C}_{3} &  & \nonumber 
\end{eqnarray}
where the conditional probabilities are given by Eq. (\ref{dist2}).

The generalization to other cases in which \( h^{C}_{i} \) is not
necessarily positive, is straightforward. We choose at random two
specific local fields for the two parties \( h^{A}_{i} \) and \( h^{B}_{i} \),
from their joint probability distribution which is derived from the
correlation matrix, making use of the overlap between the parties'
units. We then proceed to calculate the probability of each unit of
the attacker to be the one with the lowest absolute local field value,
given by Eq. (\ref{probLF1}). Once we have \( P_{i} \), \( i=1,2,3 \)
( \( P_{i} \) is the probability that unit \( i \) has the lowest
local field value), we use an auxiliary random number \( p_{\alpha } \),
to choose the unit to be flipped:\begin{equation}
\label{choosei}
\hat{\tau }_{i}=\tau _{i}\left[ 1-2\theta \left( p_{\alpha }-\sum ^{i-1}_{j=1}P_{j}\right) \theta \left( \sum ^{i}_{j=1}P_{j}-p_{\alpha }\right) \right] 
\end{equation}
where \( P_{0}=0 \).

\begin{figure}
{\centering \resizebox*{0.5\textwidth}{0.3\textheight}{\rotatebox{270}{\includegraphics{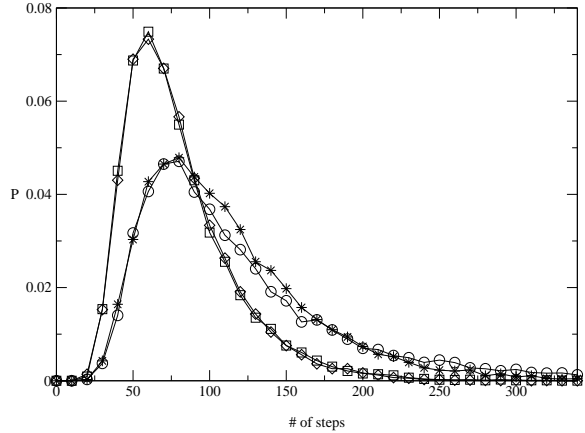}}} \par}

\caption{\label{times_dist} The synchronization time and learning time distribution
for the flipping attacker, obtained by simulations with \protect\( N=10^{3}\protect \)
(diamonds/stars for synchronization/learning) and analytical calculations
(squares/circles for synchronization/learning ) with \protect\( L=3\protect \),
averaged over \protect\( 10^{4}\protect \) runs.}
\end{figure}
\begin{figure}
{\centering \resizebox*{0.5\textwidth}{0.3\textheight}{\rotatebox{270}{\includegraphics{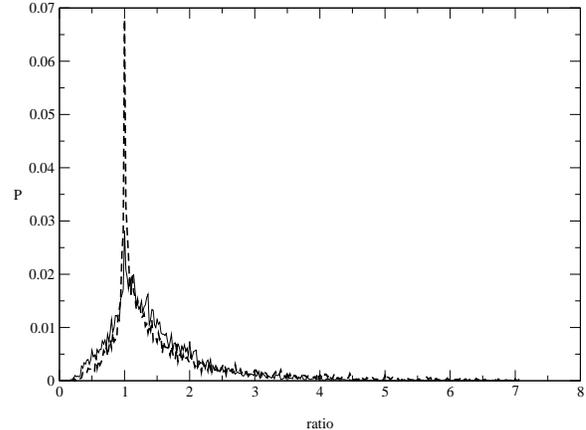}}} \par}

\caption{\label{ratio_flipp}The distribution of the ratio \protect\( R=t_{learn}/t_{synch}\protect \),
obtained by simulations (dashed line) with \protect\( N=10^{3}\protect \),
and analytical (solid line) results, with \protect\( L=3\protect \),
averaged over \protect\( 10^{4}\protect \) runs. .}
\end{figure}

Simulations and analytical calculations with \( L=3 \), \( N=10^{3} \)
averaged over \( 10^{4} \) runs, indicate that the flipping attacker
is successful. In figure \ref{times_dist} we plotted the synchronization
time and learning time distribution for the flipping attack, obtained
by simulations (circles for synchronization and squares for learning)
and analytical calculations (squares for synchronization and triangles
for learning). The flipping attacker's ability can be measured by
the ratio of the attacker learning time and the parties' synchronization
time, \( R=t_{learn}/t_{synch} \). Figure \ref{ratio_flipp} shows
the distribution of this ratio for simulations (dashed line) and analytical
(solid line) results. The probability of the flipping attacker to
finish learning before synchronization is achieved by the parties
is \( 28\% \), as presented in Figure \ref{ratio_flipp}.

\subsection{Discussion\label{DMdiscus}}

In the previous section we introduced macro-dynamical updating equations
that imitate the simulation results of discrete mutual and dynamic
learning. All numeric runs of the macro-dynamical equations are in
good agreement with simulations. The TPMs that perform mutual learning
synchronize in a finite number of steps that scales with \( \ln N \).
The macro-dynamical updating equations describe the system in the
limit of \( N\rightarrow \infty  \), and they result in an exponential
decay of the order parameter \( \rho  \) to \( -1 \), where receiving
the exact value of \( -1 \) depends on computer accuracy. However,
defining the synchronization by any finite and close to \( -1 \)
value, results in a synchronization state that is achieved in a finite
number of steps even in the thermodynamic limit. The good fit in that
limit between analytical results and simulations results is indicated
in Figures \ref{dist1}, \ref{times_dist} and \ref{ratio_flipp}.
We presented here analytical results in the case of continuous as
well as discrete weight vectors. Recently, \cite{Mislovaty} the scaling
between \( N \) and \( L \) was discussed, based on large scale
simulations with different \( L \) and \( N \) values. It may be
interesting to develop the numerical equations in the limit of infinite
\( L \) and to find the appropriate interplay between these two quantities. 

We conclude by presenting the potential of the TPMs to serve as a
public key cryptosystem. This is based upon the following features:
the synchronization state may serve as the key in a certain encryption
and decryption rule. This key evolves in public without the need of
prior communication; one needs only to perform a finite number of
steps of exchanging inputs and outputs in order to converge to a synchronized
state. The analytical derivation shows that even for infinite large
systems, \( N\rightarrow \infty  \), there will be finite distribution
of synchronization times (where synchronization time is defined by
\( \rho =-1+\epsilon  \) where small \( \epsilon  \) is a coefficient)
and the synchronization time itself will be finite. The flipping attacker
succeeds in revealing the secret for small \( L \) values, as \( L \)
enlarges the task becomes harder for her \cite{Mislovaty}. It is
yet to be determined whether it is possible to make better use of
the information in the channel, and to device a strategy that performs
perfect learning on the average in the same number of steps typical
for synchronization even for large \( L \).

\begin{acknowledgments}
I.K. acknowledges partial support of the Israel Academy of Science.
This paper is part of the Ph.D. Thesis of M.R.
\end{acknowledgments}

\end{document}